\documentclass[a4paper,11pt]{article}
\usepackage{pos}
\usepackage{comment} 
\usepackage{wrapfig}  

\title{Liquid Scintillators; Technology and Challenges}

\author*[a]{Milind V. Diwan}

\affiliation[a]{Brookhaven National Laboratory,\\
  Physics Department, Upton, New York, USA}

\emailAdd{diwan@bnl.gov}

\abstract{This is a brief review of liquid scintillators, an important technology for detection of ionizing radiation.  We will first review the basic mechanisms of light production in most organic liquid scintillators.  For most practical detector applications, the scintillators need to be optimized for  choices of photosensors and compatibility with optical windows.  A summary of important past experimental projects with liquid scintillators is provided.  We will complete the review with a list of modern practices, particularly of metal doping, and development of water based hybrid materials that allow simultaneous detection of Cherenkov and scintillation light.   }

\FullConference{6th International Conference on Technology and Instrumentation in Particle Physics (TIPP2023)\\
 4 - 8 Sep 2023\\
Cape Town, Western Cape, South Africa\\}

\begin{document}
\maketitle

\section{Introduction}

In an astonishing paper \cite{stokes1852} George Stokes explained that the nature of light is independent of its origin, and that upon what he calls "dispersion" in materials the invisible rays (ultraviolet light) shift to rays of visible light with lower "refrangibility". He also remarks that this effect is particularly common in organic liquids. In his words: "Although the passage through a thickness of fluid amounting to a small fraction of an inch is sufficient to purge the incident light from those rays
which are capable of producing epipolic dispersion, the dispersed rays themselves traverse many inches of the fluid with perfect freedom. It appears therefore that the
rays producing dispersion are in some way or other of a different nature from the dispersed rays produced."
After centuries of observing the phenomena of fluorescence and phosphorescence, we finally had a model based on science. Today the shift of wavelength between absorbed and emitted light is called the Stokes shift and is the central property that allows us to use scintillating materials for radiation detection.   

Molecular excitation in fluorescent organic liquids can be caused by either absorption of  UV photons or through the ionization of charged particles as they traverse 
the scintillation material.  Some of the de-excitation energy is lost by the emission of light through fluorescence, which does not change the electron spin state, and phosphorescence,  which changes the spin state, and therefore has a longer lifetime for the excited state. The Stokes shift between absorbed and emitted light is caused by the dynamics of the molecular quantum states, and has a generic nature that is useful to review.  

\begin{figure}[b]
    \centering
    \includegraphics[width=1.1\textwidth]{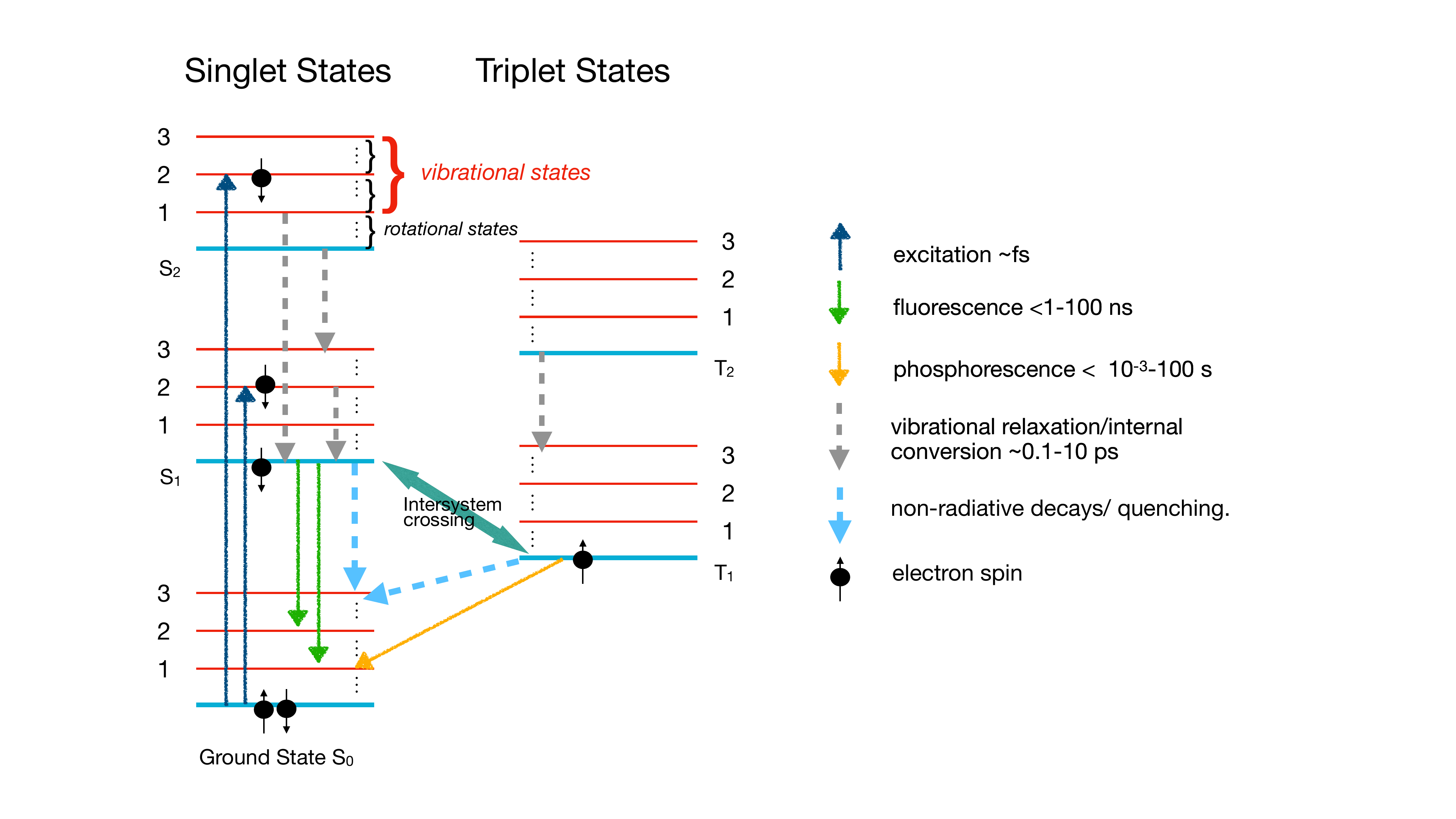}
    \caption{Jablonski diagram showing the dynamics of photon absorption and emission for common scintillation detectors.} 
    \label{fig:jablonski}
\end{figure}

Figure \ref{fig:jablonski} is a typical Jablonski diagram that describes molecular states and their excitations in a schematic format.  The quantum states of molecules can be described using the Born-Openheimer approximate method in which the electronic (due to the motion of the electrons), vibrational (due to the motion of the nuclei), and rotational states can be separated.  Typical level differences for electronic states correspond to wavelengths of $\sim 200-500$ nm;  vibrational spectroscopy corresponds to  $\sim 3 - 30 \mu$m; and rotational spectroscopy  is in the microwave range.  The electronic singlet states are labeled as $S_{0n}, ~S_{1n}, ...$, etc, where the second index is for the vibrational state. Singlet states get excited by the elevation of an electron in energy without a change in its spin; however coupling of orbital angular momentum and spin can cause triplet excitations labeled as $T_{1n}, ~T_{2n}$ etc. 
Triplet states have a  lower energy than corresponding singlet states, however the transitions are forbidden  due to angular momentum selection rules.  

Transitions between states that lead to absorption or emission of light are governed to leading order by two important principles. The Frank-Condon principle states that the electronic transitions are much faster that vibration ones, and that transitions happen between vibrational states, such as   $S_{00}$ and $S_{1n}$,  in order of their quantum wave-function overlap.  A second rule (Michael Kasha's rule) is that molecules in excited electronic states quickly relax to the lowest vibrational states before decaying to the electronic ground state via light emission.  These two rules lead to the absorption and emission spectra that are commonly seen in most scintillation materials.  The absorption spectra tend to be at higher energy (lower wavelength) because of the quantum mechanical preference of $S_{00}$ states to transition to $S_{1n}$ and the emission spectra tend to be shifted to higher wavelengths due to the preference of $S_{10}$ states to transition to $S_{0n}$. Any small overlap in the absorption and emission spectra leads to inefficiency of the photon yield (called quantum yield) from such materials.   This is the  origin of the Stokes shift that is the basis of almost all organic scitillation detector materials. 

Some of the energy loss due to ionizing radiation in organic scintillation materials leads to molecular excitations in the higher electronic and vibrational states. This process occurs very fast and populates $S_{1n}$ and $T_{1n}$ states. If the ionization density is high, then the $T$ states get relatively higher numbers. Molecules rapidly deexcite non-radiatively to the lowest vibrational states and then emit a photon to de-excite to the electronic ground state.  
The emission from the singlet, S, states tends to have faster emission rates than emission from the triplet states. This leads to distinct pulse shape differences between low and high ionization density events providing a tool for particle identification.  
A number of mechanisms govern saturation effects (Berks' effect) as well as energy transfer mechanism, such as quenching, that lead to no photon emission\cite{Bir}. 
Any good textbook or review paper on photochemistry can be used to understand further details of the basic mechanisms described above \cite{atkins,bohm}.  

\section{Application in radiation detectors }

Given the brief nature of this review, it is impossible to provide a complete description  of the use of liquid scintillators in the past.  However some notable examples provide historical markers and lessons.  A much more complete investigation can be found in reference \cite{buckyeh}.  Given the bulk nature of liquid scintillators and their inherent low cost, they have found excellent applications  in detection of neutrinos. They are also extremely valuable in total absorption calorimetry.  

Metal loaded organic liquid scintillators (LS) have been used in neutrino detectors from the first neutrino experiment of Reines and Cowan~\cite{R53}. They searched for reactor anti-neutrino interactions in a cadmium loaded LS.  This first application established the advantages of such a detector type as high purity, low energy threshold and totally active.  
Other advantages are 
detector homogeneity, flexible handling, scalability to large volumes,  and relatively low cost since most of the material is based on the petrochemical industry.    
The experiment E734 at BNL in the 1980's was a 170 ton segmented liquid scintillator target for accelerator neutrinos \cite{e734}. The scientific goal was to measure the weak mixing angle ($\sin^2 \theta_W$)  using elastic scattering of neutrinos and antineutrinos.  This experiment established the use of liquid scintillator for high resolution electromagnetic calorimetry and   precise event timing.  More recently, liquid scintillators have played the central role in establishing the final mixing angle in the neutrino sector ($\sin^2 \theta_{13}$) by the Daya Bay\cite{DB2014}, Reno\cite{RENO}, and Double Chooz \cite{DC1} experiments.  In particular, Daya Bay established three practices associated with organic  liquid scintillators:  these materials allow very precise calibration and uniformity of response across many modules and long periods of time;  stable gadolinium metal loading for neutron detection can be achieved, and extremely precise low energy calibration can be achieved for measurement of reactor spectra from inverse beta decay reactions. 

Experiments Borexino~\cite{Bor} and KamLAND~\cite{Kam} have demonstrated the capabilities of LS in experiments with very low count rates in the MeV region. Several different purification techniques are necessary  to improve the performance of a LS detector for low backgrounds. An essential step is nitrogen purging to remove spurious gases dissolved in the liquid such as oxygen, which has a quenching effect on the scintillator light yield. Highest light yields of about 13000~photons/MeV are obtained with Pseudocumene (PC) based scintillators. For Borexino, which is using pure PC as solvent, the yield is a little bit lower, 11500~photons/MeV~\cite{Eli}, since the PPO concentration is rather low (1.5~g/l). In KamLAND the PC is diluted in n-dodecane. The aromatic fraction is at 20\%, nevertheless the LY is still more than 80\% of the Borexino value at the same PPO concentration~\cite{Sue}.  Finally, 
JUNO will be a the world's largest liquid scintillator detector with 20 kton of  based on using Linear Alkyl Benzene with PPO and bis-MSB as shifters\cite{JUNO}.

\section{Common organic materials and fluors }
\begin{wrapfigure}{r}{0.5\textwidth}
\begin{center}
 \vskip -1.0cm
    \includegraphics[width=0.34\textwidth]{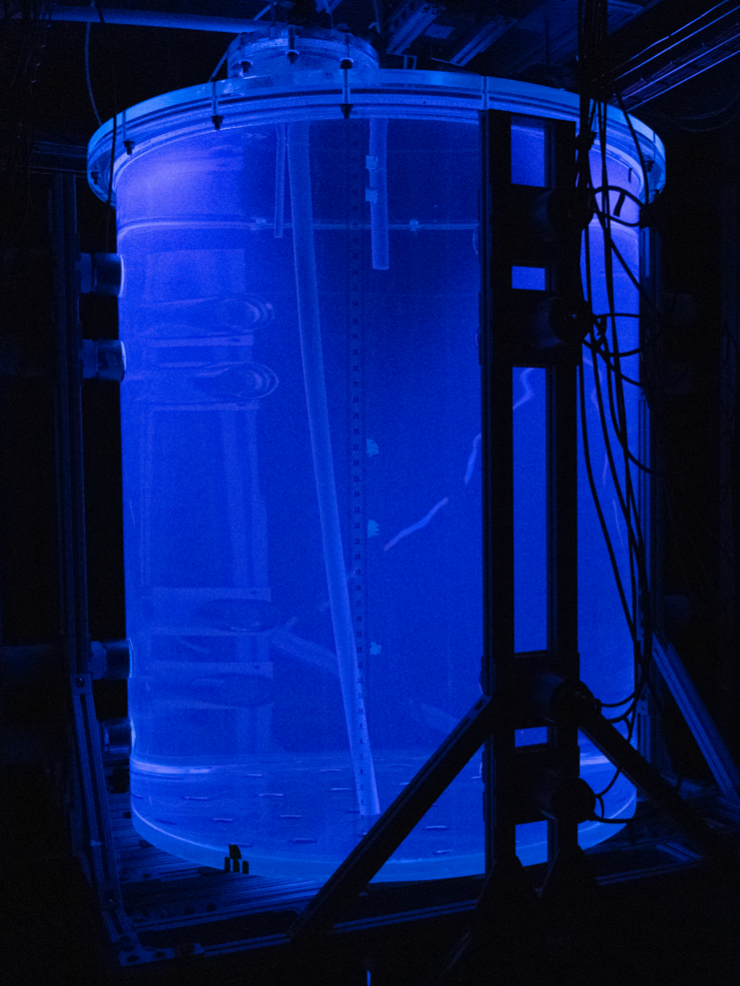} 
\end{center} 
    \caption{ The 1-ton WbLS liquid illuminated by two UV lamps positioned at the top, causing it to scintillate blue light.}
    \label{fig:detector}
    \end{wrapfigure} 
\begin{table}[h]
\caption[Solvents]{Density, flash point and the wavelengths of the optical absorption/emission peaks (dissolved in cyclohexane) for several solvent candidates are shown\cite{buckyeh}. \label{solvents}}
\begin{center}
\begin{tabular}{lccrcc}
Molecule & chemical formula & density [kg/l] & flash point & abs.~max. & em.~max. \\
\hline
PC & C$_9$H$_{12}$ & 0.88 & 48$^\circ$C & 267~nm & 290~nm \\
toluene & C$_7$H$_8$ & 0.87 & 4$^\circ$C & 262~nm & 290~nm\\
anisole & C$_7$H$_8$O & 0.99 & 43$^\circ$C & 271~nm & 293~nm\\
LAB & -- & 0.87 & $\sim140^\circ$C & 260~nm & 284~nm \\
DIN & C$_{16}$H$_{20}$ & 0.96 & $>140^\circ$C & 279~nm & 338~nm \\ 
o-PXE & C$_{16}$H$_{18}$ & 0.99 & 167$^\circ$C & 269~nm & 290~nm \\
n-dodecane & C$_{12}$H$_{26}$ & 0.75 & 71$^\circ$C & -- & -- \\
mineral oil & -- & 0.82 -- 0.88 & $>130^\circ$C & -- & -- \\
\hline
\end{tabular}
\end{center}
\end{table}
\begin{table}[h]
\caption[Fluors]{The wavelengths of the optical absorption and emission peak maxima for primary and secondary fluor molecules diluted in cyclohexane are listed\cite{buckyeh}. \label{fluors}}
\begin{center}
\begin{tabular}{llcccc}
Molecule & chemical formula & abs.~max. & em.~max.\\
\hline
PPO	& C$_{15}$H$_{11}$NO & 303~nm & 358~nm \\ 
PBD & C$_{20}$H$_{14}$N$_2$O & 302~nm & 358~nm \\
butyl-PBD & C$_{24}$H$_{22}$N$_2$O & 302~nm & 361~nm \\ 
BPO & C$_{21}$H$_{15}$NO & 320~nm & 384~nm \\
p-TP & C$_{18}$H$_{14}$ & 276~nm & 338~nm \\
TPB & C$_{28}$H$_{22}$ & 347~nm & 455~nm\\
bis-MSB & C$_{24}$H$_{22}$ & 345~nm & 418~nm \\
POPOP & C$_{24}$H$_{16}$N$_2$O$_2$ & 360~nm & 411~nm \\
PMP & C$_{18}$H$_{20}$N$_2$ & 295~nm & 425~nm \\
\hline
\end{tabular}
\end{center}
\end{table}
Tables \ref{solvents} and \ref{fluors}  provide a list of key ingredients for modern organic liquid scintillator formulations; the full form of the chemicals can be obtained from \cite{buckyeh}.  The selection of solvents -- which are mostly variants of benzene -- and wavelength shifters follow the most important requirements from specific experiments: these are light yield,  light emission spectrum, decay times, optical clarity or attenuation, and radiopurity.  Engineering issues are also important such as compatibility with common container and window materials, and flash point temperature.  
A primary limitation is due to wavelength sensitivity of common photo-sensors. A typical bialkali photomultiplier is sensitive in the range of  350 nm  to 600 nm \cite{hpk}.   With high grade fused silica windows, the sensitivity can be brought down to $\sim$200 nm, but these are only useful in specialized applications. New silicon photosensors (SiPM or MPPC) \cite{hpk}  are also becoming important, and they can be configured for vacuum ultraviolet  wavelengths, but these are still small in dimension ($\sim 5$ mm) and have high noise at room temperature.  Any photonic emission in the liquid therefore must be shifted to the wavelength matching appropriate  sensors -- mostly above $\sim$ 350 nm.  
This shifting is performed by introducing one or two fluors that absorb the ionization energy from the primary solvent through the Forster mechanism and emitting it at the desired wavelength\cite{Fo48}.

\section{Modern developments}

As remarked earlier,  bulk liquid scintillators have become extremely important for neutrino detection. They provide three primary advantages:  the LS acts as a totally active  target for neutrinos, the presence of free protons gives excellent sensitivity for reactor anti-neutrinos, and the ability to tailor the material with fluors gives high light yield and flexibility in coupling to appropriate photo-sensors.   

In neutrino physics it is quite desirable to introduce metals in the liquid. This is either for absorption and detection of neutrons on gadolinium or lithium or for sensitivity to specific physics modes such as neutrino less double beta candidates.  
Last two decades has seen great  improvement in metal loading techniques in organic liquids. 
The most promising procedure for metal loading is the preparation of an organo-metallic complex that is soluble in liquid scintillator. Modern experiments (such as Daya Bay) have used metal corboxylates for this purpose and have obtained excellent attenuation lengths ($\sim$ 20 m) and light yield  \cite{Yeh2011}.

\subsubsection{Water Based Liquid Scintillator}  

Generally it is  easier  to add metals to aqueous solutions which can be mixed into an organic liquid with surface active agents (surfactants) which have both hydrophilic and hydrophobic chemical groups.  
A direct growth out of the effort of metal loading is the recent interest and development of water based liquid scintillators\cite{WLS11, WLS15a}. WbLS is now being investigated at a scale of several tons \cite{onetonpaper, eos} with the aim towards a large underground detector \cite{Theia}. WbLS is a cost-effective medium that could enhance future massive Cherenkov detectors with the unique capability of detection below Cherenkov threshold.  Such a hybrid technique has been demonstrated by a number of experiments \cite{BOREXINO:2021xzc, SNO:2023cnz}.  The new WbLS formulation used in the  1-ton and 30 ton systems at BNL consists of a mix of modified surfactants, a fluor of 2,5-diphenyloxazole (PPO), and a DIN-based (di-isopropylnaphthalene) liquid scintillator, all commercially available. All the raw components were purified by in-house thin-film vacuum distillation and recrystallization. A newly developed in-situ WbLS mixing technique largely simplifies the process by first filling the detector with high purity water, separately manufacturing the organic scintillator,  and then sequentially injecting into the vessel allowing WbLS mixing in-situ through the operating circulation system. This process was shown to be effective within a reasonable timescale yielding the expected non-Cherenkov light yield of about $128\pm 26$ photons emitted per MeV for 1\% by mass mixture of LS with water (see figure \ref{fig:detector}). The Cherenkov light yield is consistent with pure water. Further work is in progress to understand the photo-chemistry in this novel medium and limit the statistics  of absorbed and re-emitted Cherenkov light.  Loading of the material with gadolinium, and associated purification and separation systems are being developed.    

\section{Conclusion and Future Challenges}

Liquid scintillator technology and applications,  particularly for neutrino physics, have reached new levels of scale and performance.  Noble liquid detectors -- not covered in this review --  are also getting scaled to extraordinary sizes.  The DUNE liquid argon detector, the Darkside-20, and LZ and XenonNT projects with liquid xenon  will have unprecedented scale and sensitivity.  The JUNO project will be the largest organic liquid scintillator detector for a long time. Despite these astonishing scales, there are deep and fundamental  
challenges in understanding and utilizing of these scintillators.   
Detailed understanding of molecular states for individual organic molecules (as well as noble liquid dimers) can be obtained in gas states, but the shifting of levels and transitions in coupled liquids is  understood only qualitatively. Energy loss mechanisms for ionizing particles in liquids can also be modelled only qualitatively. For example,  in most experiments the Bethe-Bloch  formula and the Landau-Vavilov theory is mostly used with adjustments using ad-hoc calibration procedures.  A deep challenge would be to devise a simulation of energy loss with detailed understanding of cross sections and molecular transitions in at least some of the most common scintillation media (e.g. see \cite{dpb2015}). Software tools with sufficient microscopic detailed description of coupled optical systems, fluors, windows, and detection capabilities could significantly aid in optimized detector designs.  
Such tools could be built on modern GPU based software architectures \cite{Althueser2022}. 
There are also challenges on the detection instrumentation side. New optical imaging technologies are emerging. These include dichroic filters \cite{eos} and conformal coatings with adjustable parameters and advanced single photon avalanche diode (SPAD) based cameras with low noise performance. Such technologies also need development of low cost VUV capable optical windows in a variety of environments (cryogenic and liquid based).  New advanced software simulation programs and advancements in materials promises to open new avenues of liquid scintillator technologies for applications in particle and nuclear physics.

\bibliographystyle{unsrt}
\bibliographystyle{plain} 


\end{document}